\documentclass[twocolumn,oldversion]{aa}
\usepackage{graphicx}
\usepackage{txfonts}
\input{psfig.sty}
\begin{document}

\title{``Expansion" around the vacuum: how far can we go from $\Lambda$?}


   \author{
          J. S. Alcaniz \inst{1}
          \and
          H. \v Stefan\v ci\'c\inst{2,} 
          \thanks{On leave of absence from Theoretical Physics Division, Rudjer Bo\v{s}kovi\'{c} Institute, Zagreb, Croatia}  }

   \offprints{J. S. Alcaniz}

   \institute{Observat\'orio Nacional, Rua Gal. Jos\'e Cristino 77, 20921-400 Rio de Janeiro - RJ,
Brasil\\
\email{alcaniz@on.br}
         \and 
Departament d'Estructura i Constituents de la Mat\`{e}ria, Universitat de Barcelona, Av. Diagonal 647, 08028 Barcelona, Catalonia, Spain\\ \email{stefancic@ecm.ub.es}            }
      

   \date{Received ; accepted}

\abstract{The cosmological constant ($\Lambda$), i.e., the energy density stored in the true vacuum state of all existing fields in the Universe, is the simplest and the most natural possibility to describe the current cosmic acceleration. However, despite its observational successes, such a possibility exacerbates the well known $\Lambda$ problem, requiring a natural explanation for its small, but nonzero, value. In this paper we discuss how different our Universe may be from the $\Lambda$CDM model by studying observational aspects of a kind of ``expansion" around the vacuum given by the equation of (EOS) $p_{d}=-\rho_{d} - A \rho_{d}^{\alpha}$. In different parameter regimes such a parametrization is capable of describing both quintessence-like and phantom-like dark energy, transient acceleration, and various (non)singular possibilities for the final destiny of the Universe, including singularities at finite values of the scale factor, the so-called ``Big Rip", as well as sudden future singularities. By using some of the most recent cosmological observations we show that if the functional form of the dark energy EOS has additional parameters very little can be said about their values from the current observational results,  which postpones, until the arrival of more precise observational data, a definitive answer to the question posed above.

   }

   \maketitle
%

\section{Introduction}

The understanding of the global evolution of the observationally amenable universe, mathematically encoded in the dynamics of its scale factor $a$, is of utmost importance in explaining practically all cosmological phenomena. One of the most intriguing aspects of this evolution is the recently established late-time transition from a decelerated to an accelerating regime of the expansion of the Universe. The evidence confirming the low-redshift onset of the accelerating expansion phase comes directly from current type Ia supernovae (SNIa) measurements (Riess et al. 1998; 2004; Perlmutter et al. 1999), and is indirectly supported by various other cosmological observations such as the measurements of the cosmic microwave background (CMB) anisotropies (de Bernardis et al. 2000; Spergel et al. 2003; 2006), and the current large scale structure (LSS) data (Tegmark et al. 2004). 

Although the actual nature of the mechanism behind the current cosmic acceleration, as well as its link with fundamental interactions, constitutes nowadays a completely open question, there are many, often very ingenious, attemps of modeling the current stage of the cosmic evolution. Some of the most important questions that these models aim at answering, and with respect to which different scenarios may be distinguished, are:

\begin{enumerate}

\item {\em What is the mechanism causing the accelerating expansion of the universe?}

The idea of a late time accelerating  universe is usually associated with unknown physical processes involving either new fields in high energy physics or modifications of gravity at very large scales. In following the former route, the predominant concept nowadays is the existence of a new component with a sufficiently negative presure, named {\em dark energy} (Sahni \& Starobinsky 2000; Peebles \& Ratra 2003; Padmanabhan 2003), which is fully characterized by its equation of state (EOS), $w(t) = p_{d}(t)/\rho_{d}(t)$, where $p_{d}(t)$ and $\rho_{d}(t)$ are, respectively, the dark component pressure and energy density. The value $w=-1$ characterizes the {\em vacuum energy} ($\Lambda$), which is conceptually the simplest model ($\Lambda$CDM). Although such scenarios constitute a kind of benchmark model in the analysis of observational data, the large discrepancy between the theoretically predicted and ``observed" values of $\Lambda$ has incited the study of dynamical dark energy (DDE) models. 

In this latter case, $\Lambda$ is treated as a dynamical quantity whereas its constant EOS, $w=-1$, is preserved. This comprises, among others, models based on renormalization group running $\Lambda$ (Shapiro \& Sola, 2000; 2002;  Babic et al. 2002; Shapiro, Sola \& Stefancic, 2005; Sola \& Stefancic, 2005) and vacuum decay ($\ddot{\rm{O}}$zer \& Taha, 1986; Bertolami, 1986; Cunha, Lima \& Alcaniz, 2002; Alcaniz \& Lima, 2005). Other DDE models, in which the potential energy density associated with a dynamical scalar field ($\phi$) dominates the dynamics of the low-redshift Universe, have also been extensively discussed in the current literature (see, e.g., Ratra \& Peebles, 1988; Wetterich, 1988; Caldwell, Dave \& Steinhardt, 1998). The EOS parameter in this class of models is necessarily  a function of time and may take values $> -1$ [{\em quintessence}] or $< -1$ [{\emph{phantom}}] (Caldwell, 2002; Faraoni, 2002; Alcaniz, 2004). In this regard, it is still worth mentioning that some recent observational analyses (Melchiorri et al. 2003), implying that a transition from
$w>-1$ to $w<-1$ might have happened at low redshifts, have also instigated the development of models with $\Lambda$ boundary crossing (Nojiri \& Odintsov, 2005; Feng, Wang \& Zhang, 2005; Stefancic, 2005). 

The second route to the acceleration problem involves modifications of gravity at cosmological scales (without dark energy) (Randall \& Sundrum, 1999; Dvali, Gabadadze \& Porrati, 2000; Maia et al. 2005). An interesting example of this approach are the so-called braneworld models, which offer an intriguing mechanism for the acceleration of the universe and have been sucessfully tested by a number of observational analyses (Alcaniz, 2002; Jain, Dev \& Alcaniz, 2002; Lue 2003).

\item {\em What is the duration of the accelerating expansion phase?}

Although the transition from an initially decelerated to a late-time accelerating expansion is becoming observationally well established, the acceleration redshift $z_{acc}$, as well as the duration of the accelerating phase, depends crucially on the cosmological scenario [see, e.g., Bassett, Corasaniti \& Kunz 2004]. Several models imply an eternal acceleration (e.g. $\Lambda$CDM model) or even an accelerating  expansion until the onset of a cosmic singularity (e.g., phantom models with constant EOS). However, an eternally accelerating universe seems not to be in agreement with String/M-theory since it is endowed with a cosmological event horizon which prevents the construction of a conventional S-matrix describing particle interactions within the framework of these theories (Fischler et al. 2001) (see also Ellis, Mavromatos \& Nanopoulos, 2005). In that sense, cosmological models that predict the possibility of a transient acceleration phenomenon are of special interest.

\item {\em What is the fate of the presently accelerating Universe?}

As well known, information on the fate of the Universe is encoded in the form of its past expansion and depends fundamentally on the nature of its dominant energy component. As an example, the so-called phantom energy models ($w < -1$) predict a very dramatic fate for the Universe, i.e., a universe which ends with a future singularity, the so-called {\em Big Rip} (Caldwell, Kamionkowski and Weinberg 2003; Nesseris \& Perivolaropoulos, 2004). As shown by Lima \& Alcaniz (2004), the thermodynamic fate of a dark energy dominated universe may also be considerably modified, with the possibility of a increasingly hot expanding universe.

\end{enumerate}

In this paper we aim at addressing some of the above questions in the context of a dark energy model characterized by the following EOS (Stefancic, 2005; Nojiri \& Odintsov, 2004)
\begin{equation}
\label{eq:oureos}
p_{d} = -\rho_{d} - A \rho_{d}^{\alpha} \, .
\end{equation}
In this simple framework, by using the current observational data to place limits on the parameters of (\ref{eq:oureos}), one can simultaneously test if the nature of the dark energy is quintessence-like ($A < 0$) or phantom-like ($A > 0$), if the acceleration is eternal or transient, and which form of future asymptotic (non)singular behavior is implied observationally. The organization of the paper is the following: Section II presents a theoretical analysis and discusses the most interesting phenomenological implications of the model. Section III is devoted to the comparison of the model against the observational data. We end this paper by summarizing our main results in the conclusion Section.

\section{The model}

The EOS (\ref{eq:oureos}) can be understood as  a sort of ``expansion" around the vacuum EOS. The \emph{a priori} motivation for the study of such a dark energy EOS is twofold. First, the $\mathrm \Lambda$CDM cosmology, characterized by the ``true" cosmological constant, is the standard in the analysis of the observational data. This means that the measurement of a possible deviation of the dark energy EOS from the vacuum EOS is presently one of the highest priorities in Cosmology. The model (\ref{eq:oureos}) provides a suitable framework for such a study. Second, from the theoretical side, the model (\ref{eq:oureos}) represents a dynamical dark energy model with a time-dependent EOS parameter, which can describe both quintessence and phantom types of dark energy behavior. A detailed analysis of the dynamics of the universe with a dark energy component (\ref{eq:oureos}) was given by Stefancic (2005). In what follows we summarize the main results of this analysis.

The conservation of the energy-momentum tensor for the dark energy model (\ref{eq:oureos}) yields the scaling of dark energy density with the scale factor of the universe, i.e., 
\begin{equation}
\label{eq:scalour}
\rho_{d}=\rho_{d,0} \left( 1 + 3 \tilde{A} (1-\alpha) \ln a \right)^{1/(1-\alpha)} \, ,
\end{equation}
where we have set $a_0 =1$ and the abbreviation $\tilde{A}=A \rho_{d,0}^{\alpha-1}$ was introduced. The parameter $w=p_{d}/\rho_{d}$ of the EOS (\ref{eq:oureos}) is given by
\begin{equation}
\label{eq:w}
w = -1 - \frac{\tilde{A}}{1 + 3 \tilde{A} (1 - \alpha) \ln a} \, .
\end{equation}
This equation reveals that for different signs of $\tilde{A}$ we may expect different behavior of the dark energy component, which presently generates the acceleration of the Universe. For example, for $\tilde{A} > 0$, $w$ remains below -1 during the entire evolution of the universe, whereas for $\tilde{A} < 0$ the EOS parameter remains always above the $\Lambda$ barrier ($w = -1$) and may become even  larger than $-1/3$, which means that the dark energy component can change its action from accelerating to decelerating.
The expression for $p_d = w \rho_{d}$ can be obtained directly from Eqs. 
(\ref{eq:scalour}) and (\ref{eq:w}) in a straightforward manner. 

Here we assume that the universe contains two components, namely, the dark energy component and the nonrelativistic matter component. Thus, the Friedmann equation is given by
\begin{equation}
\label{eq:Fried}
\left( \frac{\dot{a}}{a} \right)^2 + \frac{k}{a^2} = \frac{8 \pi G}{3} (\rho_{d}+\rho_{m}) \, .
\end{equation}

In the case when the dark energy density dominates over the energy density of nonrelativistic matter and the spatial curvature term $k/a^2$, Eq. (\ref{eq:Fried}) can be solved analitically to obtain the asymptotic time-dependence of the scale factor $a$. For $\alpha \neq 1/2$ the time-dependence of  the scale factor is given by 
\begin{eqnarray}
\label{eq:aodt}
& &\left( 1 + 3 \tilde{A} (1-\alpha) \ln a_{1} \right)^{(1-2\alpha)/(2(1-\alpha))} \nonumber \\
&-& \left( 1 + 3 \tilde{A} (1-\alpha) \ln a_{2} \right)^{(1-2\alpha)/(2(1-\alpha))} \nonumber \\
&=&\frac{3}{2} \tilde{A} (1-2\alpha) \Omega_{d,0}^{1/2} H_{0} (t_{1}-t_{2}) \, ,
\end{eqnarray}
whereas for $\alpha=1/2$ one obtains 
\begin{equation}
\label{eq:aodtspec}
\ln \frac{1 + \frac{3}{2} \tilde{A} \ln a_{1}}
{1 + \frac{3}{2} \tilde{A} \ln a_{2}} = \frac{3}{2} \tilde{A} \Omega_{d,0}^{1/2} H_{0} (t_{1}-t_{2}) \, .
\end{equation}
In the expressions given above we have introduced the notation  $\Omega_{d,0}=\rho_{d,0}/\rho_{c,0}$, where $\rho_{c,0} = 3 H_{0}^2/(8 \pi G)$ is the present critical energy density. Throughout this paper the subscript $0$ refers to the present epoch of the evolution of the Universe.

Using equations (\ref{eq:scalour}), (\ref{eq:aodt}), and (\ref{eq:aodtspec}) it is possible to determine the characteristic values of the model parameters which separate qualitatively different behaviors of the model. The characteristic value of the parameter $\tilde{A}$ is $\tilde{A}=0$. Thus, for $\tilde{A} > 0$ the model has a phantom-like behavior, for $\tilde{A} < 0$ the model behaves as quintessence, while for $\tilde{A}=0$ the model is equivalent to the $\Lambda$CDM scenario. The characteristic values for the second parameter $\alpha$ are $\alpha=1$ and $\alpha=1/2$. The behavior of the model in the intervals delimited by these two values is described below.
The description of the rich variety of phenomena realized for different values of the parameters $\tilde{A}$ and $\alpha$ is divided into cases $\tilde{A} > 0$ and $\tilde{A} < 0$.

\subsection{$\tilde{A} > 0$}

\begin{enumerate}

\item For $\alpha > 1$ the dark energy density ($\rho_{d}$) and pressure ($p_{d}$) diverge at a {\em finite value of the scale factor}, which is also reached in a finite time interval from today. This type of singular ending of the universe, first introduced by (Stefancic, 2005; Nojiri \& Odintsov, 2004), represents a kind of singularity even stronger than the so-called ``Big Rip" singularity\footnote{For a detailed classification of singularities in phantom cosmologies, see, e.g., Nojiri, Odintsov and Tsujikawa (2005).}.

\item The case $\alpha=1$ corresponds to the well studied phantom energy with a constant EOS parameter, which leads to the well known ``Big Rip" event (Caldwell, Kamionkowski and Weinberg, 2003; Nesseris and Perivolaropoulos, 2004).

\item In the interval $1/2 <  \alpha < 1$ the scale factor $a$ diverges at a finite time, as do the dark energy density $\rho_{d}$ and pressure $p_{d}$. The value of the EOS parameter $w$ remains bounded. The singular fate of the universe in this parameter range is qualitatively similar to the ``Big Rip" singularity (Caldwell, Kamionkowski and Weinberg, 2003; Nesseris and Perivolaropoulos, 2004).

\item For $\alpha \le 1/2$ the future expansion is nonsingular and the scale factor $a$ reaches infinity in a infinite future.

\end{enumerate}

\subsection{$\tilde{A} < 0$}

\begin{enumerate}

\item The case $\alpha > 1$ is characterized by the nonsingular future expansion. The dark energy density is in this case a decreasing function of the scale factor (the cosmic time) and vanishes for $t \rightarrow \infty$.

\item The case $\alpha=1$ corresponds to the case of
quintessence-like dark energy with a constant EOS parameter $w
= -1-\tilde{A}$.

\item For $\alpha < 1$ the dark energy density reaches 0 at a
finite value of the scale factor $a_{\mathrm{NULL}} > a_{0}$ (and
at a finite time). At this scale factor value the Hubble parameter
$H$ and the total energy density are constant whereas the
parameter of dark energy EOS diverges, $w \rightarrow +\infty$.
The behavior of the dark energy pressure and the acceleration of
the universe differs for cases $0 < \alpha < 1$, $\alpha=0$, and
$\alpha < 0$.

\item[3.1.] In the case when $ 0 < \alpha < 1$, the dark energy pressure
vanishes at $a_{\mathrm{NULL}}$. For $a>a_{\mathrm{NULL}}$ the dark
energy pressure can no longer be properly defined by
(\ref{eq:oureos}). Since at $a_{\mathrm{NULL}}$ $H$ is finite, the
question is how the expansion is continued. One possibility is
that for $a>a_{\mathrm{NULL}}$ the universe expands with
$\rho_{d}=0$ and $p_{d}=0$, which also satisfies  (\ref{eq:oureos}).

\item[3.2.] For $\alpha = 0$ the dark energy density changes its sign at
$a_{\mathrm{NULL}}$ and for $a>a_{\mathrm{NULL}}$ the dark energy
pressure is positive. The expansion continues until the Hubble
parameter vanishes at some finite $a_{\mathrm{STOP}} >a_{\mathrm{NULL}}$.

\item[3.3.] For $\alpha < 0$ the dark energy pressure diverges at
$a_{\mathrm{NULL}}$, $p_{d} \rightarrow + \infty$ and $\ddot{a}
\rightarrow - \infty$. This is a concrete realization of a sudden
future singularity recently introduced by Barrow (Barrow, 2004).

\end{enumerate}

The behavior of the model for $\tilde{A} < 0$ and $\alpha < 1$ is
characterized by the finite duration of the phase of the
accelerated expansion of the universe (see Fig. 1).
The transition from the decelerated to the accelerated expansion
of the universe is the result of an interplay of the evolution of
the dark energy density and the energy density of nonrelativistic
matter. The exit from the accelerated phase is, however, the
consequence of the intrinsic dynamics of the dark energy model
(\ref{eq:oureos}) in this parameter regime. Finally, a scenario in
which the phase of accelerated expansion starts and ends in the
recent past is also allowed in this parameter regime. This
nonstandard possibility, closely related to the condition
(\ref{eq:qle0}), is also discussed in our observational analysis.

\section{Observational aspects}

The description of the model (\ref{eq:oureos}) discussed in the
previous Section clearly shows that it comprises a multitude of
cosmological solutions. In a model with such a wealth of different
possibilities constraints on the parameter space arising from
current observational data are likely to rule out many of the
possible scenarios for the evolution/fate of the Universe. This Section investigates such observational constraints by studying (i) the influence of the parameters $\tilde{A}$ and $\alpha$
on the epoch of the deceleration/acceleration (D/A) transition, as
well as on the duration of the accelerating phase and (ii) by placing observational bounds on the parametric space $\tilde{A}-\alpha$
from a statistical analysis involving a large
set of cosmological observations.

\begin{figure}[t]
\label{fig:trans}
\centerline{\psfig{figure=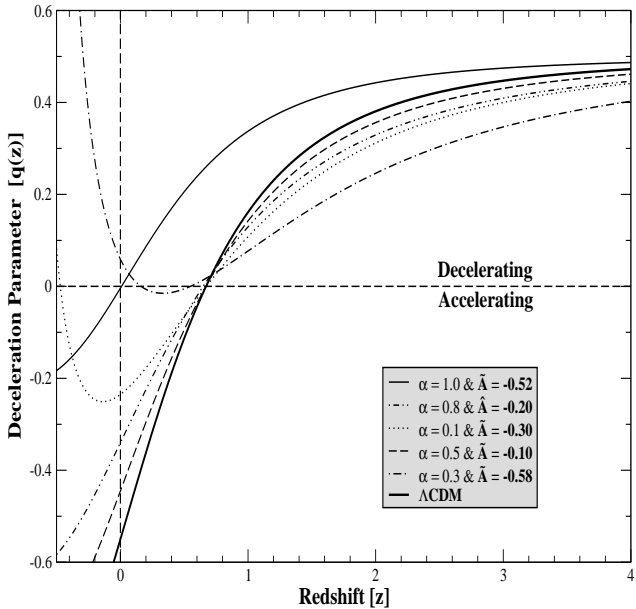,width=3.4truein,height=3.4truein,angle=-90}
\hskip 0.1in} 
\caption{The deceleration parameter as a function of redshift for some selected combinations of the pair ($\tilde{A}$, $\alpha$) and $\Omega_{m,0} = 0.3$. The horizontal dashed line divides decelerating and accelerating universes. Note that a number of different possibilities for the recent past and future of the Universe can be achieved from parametrization (1). The standard $\Lambda$CDM model (thick line) is also shown for the sake of comparison.}
\end{figure}

\begin{figure}[t]
\centerline{\psfig{figure=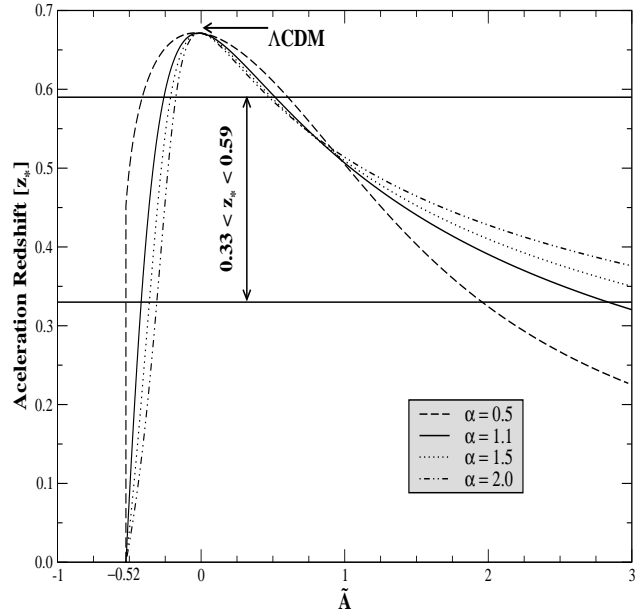,width=3.4truein,height=3.4truein,angle=-90}
\hskip 0.1in} 
\caption{The transition redshift $z_*$ as a function of the parameter $\tilde{A}$ for values of $\alpha$ lying in the interval [$\frac{1}{2}$,2]. Horizontal lines stand for the 1$\sigma$ interval $0.33 \leq z_* \leq 0.59$, as provided by current SNe Ia observations
(Riess et al., 2004). Note that our current standard model, i.e., the $\Lambda$CDM scenario predicts a transition epoch $\simeq 2\sigma$ off from the central value obtained in Riess et al. (2004).}
\end{figure}

\subsection{D/A Transition}

An important characteristic of the current SNe Ia data is the
evidence for a possible D/A transition at $z_{*} \sim 0.5$
(Riess et al., 2004). If confirmed by the upcoming data, the
precise value of $z_{*}$ will become an important observational constraint to
decisively discriminate among competing cosmological models. In
order to study the bounds on the parameters $\tilde{A}$ and
$\alpha$ from the current estimates of $z_{*}$ we first derive the
deceleration parameter as a function of the scale factor $a$, i.e.,
\begin{equation}
\label{eq:decel} q(a) = -1  + \frac{3}{2} \frac{
{\Omega_{m,0} a^{-3}} - \Omega_{d,0} \tilde{A} F(a,\alpha,\tilde{A})^{\alpha /
(1-\alpha)}}{{\Omega_{m,0} a^{-3}} + \Omega_{d,0} F(a,\alpha,\tilde{A})^{1 /
(1-\alpha)} } \, ,
\end{equation}
where $\Omega_{m,0}=\rho_{m,0}/\rho_{c,0}$ and $F(a,\alpha,\tilde{A})=1 +
3\tilde{A}(1-\alpha) \ln (a)$. As one
may check, an immediate constraint on the model parameters arises
by combining Eq. (\ref{eq:decel}) with the fact that the universe
is currently accelerated [$q(a=a_0=1) < 0$], i.e.,
\begin{equation}
\label{eq:qle0} \tilde{A} > \frac{\Omega_{m,0} - 2/3}{1 -
\Omega_{m,0}} \, .
\end{equation}
Note that the above expression does not depend on the parameter $\alpha$, a feature of the model
which will be reencountered in the statistical analysis performed
in the next section. This means that for values of $\Omega_{m,0} \simeq
0.3$, as indicated by dynamical estimates at scales up to about
$2h^{-1}$ Mpc (Calberg et al., 1996) [and also confirmed by current CMB
measurements (Spergel et al. 2004)] we find that the parameters $\tilde{A}$ is restricted to the interval  $\tilde{A} > -0.52$. Examples of the behavior of the deceleration parameter as a function of redshift are shown in Fig. 1 for selected values of $\tilde{A}$ and $\alpha$ and $\Omega_{m,0} = 0.3$. As expected from Eq. (9), irrespective of the value of $\alpha$ models with $\tilde{A} > (<) -0.52$ are currently accelerated (decelerated) while for the critical case $\tilde{A} = -0.52$ the Universe expands presently with a constant Hubble flow ($q_0 = 0$) and constitues new examples of coasting cosmologies (Kolb, 1989). As mentioned earlier, note still that the possibility of a presently decelerated universe that experienced a recent accelerated phase is also contained in the parametrization (\ref{eq:oureos}). In the particular example shown in Fig. 1 ($\tilde{A} = -0.58$ and $\alpha = 0.3$) the Universe would have entered an accelerated regime at $z \simeq 0.55$, remained accelerated for $ 2.04 h^{-1}$ Gyr, and switched again to a new decelerated period at $z \simeq 0.17$.

Another interesting behavior which is also achieved from some combinations of the pair ($\tilde{A}$, $\alpha$) has to do with the intriguing possibility that the current cosmic acceleration may not be a lasting feature. As can also be seen from Fig. 1, some scenarios of type (\ref{eq:oureos}) predict a universe that was decelerated in the past, began to accelerate at $z_* \simeq 0.5$, is currently accelerated but will eventually decelerate in the future. Such a behavior, according to some authors (Fischler et al. 2001), seems to be in full agreement with the requirements of String or M-theory, in that eternally accelerating universes are endowed with a cosmological event horizon which prevents the construction of a conventional S-matrix describing particle interactions within these frameworks\footnote{Another interesting example of transient acceleration is provided by the braneworld scenarios discussed by Sahni and Shtanov (2002).}. A typical example of an eternally accelerating universe, i.e., the $\Lambda$CDM model (thick line), is also shown in Fig. 1 for the sake of comparison.

Figure 2 shows the transition redshift $z_*$ as a function of the parameter $\tilde{A}$. Note that the behavior of $z_*$ can be easily obtained from Eq. (\ref{eq:decel}) or, more specifically, from the expression
\begin{equation}
\frac{1}{2} \Omega_{m,0} (1 + z_*)^3 - \Omega_{d,0} {\cal{F}}^{\alpha /
(1-\alpha)}\left[\frac{3}{2} \tilde{A} + {\cal{F}} \right]= 0,
\end{equation}
where ${\cal{F}} = 1 - 3\tilde{A}(1 -\alpha) \ln (1 + z_*)$. The horizontal solid lines stand for the interval $0.33 \leq z_* \leq 0.59$, which corresponds to $\pm 1\sigma$ of the estimate for the transition redshift given by Riess et al. (2004). For values of $\alpha$ lying in the interval [$\frac{1}{2}$,2] we find two branches compatible with the current estimate for $z_*$, i.e., $-0.52 < \tilde{A} \leq -0.17$ and $0.46 \leq \tilde{A} \leq 4.35$. In agreement with recent results, we also found that the $\Lambda$CDM scenario ($\tilde{A} = 0$) is $\simeq 2\sigma$ off from the central value of the current estimate for the redshift of the D/A transition.

\subsection{Statistical Analysis}

As has been mentioned earlier, the best  way to check the viability of parametrization (1) and, based on it, to discuss how different our Universe may be from the standard $\Lambda$CDM scenario, is to perform a statistical analysis involving the currently available sets of cosmological observations. To this end, we use here some of the most recent observational 
data as, for instance, the Supernova Legacy Survey (SNLS) set of 115 SNe Ia, recently published by Astier et al. (2006). These distance measurements are combined with the growth of structure index and current estimates of the age of the Universe. For the LSS, we use the the linear growth rate $f(z_{2df}) = 0.58 \pm 0.11$, measured by the 2 degree field (2dF) galaxy redshift survey (2dFGRS)(Verde et al., 2002; Lahav et al., 2002; Hawkins et al., 2003), where $z_{2df} = 0.15$ is the effective redshift of this survey and $f =d\ln{D}/d\ln{a}$ is determined by solving the equation for the linear growth rate (Lahav et al., 1991)
\begin{equation}
D^{''} + 2E(z)\,D^{'} -{3\over 2}\Omega_{m,0}(1
+z)^{3}\, D = 0,
\label{D}
\end{equation}
and primes denote $d/d(H_o t)$. The age of the Universe is assumed to be  $t^{obs}_o = 13.6 \pm 0.2$ Gyr, as obtained by MacTavish {\it et al.} (2005) from a joint analysis involving recent LSS data (the matter power spectra from the 2dFGRS and SDSS redshift surveys) and results of the most recent CMB experiments (WMAP, DASI, VSA, ACBAR, MAXIMA, CBI and BOOMERANG) [for more details on statistical analyses we refer the reader to Lima, Cunha and Alcaniz, 2003; Choudhury and Padmanabhan, 2005; Nesseris and Perivolaropoulos, 2004; Capozziello et al., 2005; Alam and Sahni, 2005.]\footnote{Note that in the analysis of the observational data one needs to take into account the peculiarities of the behaviour of the dark energy density (\ref{eq:oureos}) at positive redshifts. Namely, when $\tilde{A} (1-\alpha) > 0$, for a sufficiently high redshift the dark energy density is no longer a well defined function of redshift. A possible interpretation in this case is that the dark energy EOS (\ref{eq:oureos}) is just an effective description valid at lower redshifts and that at higher redshifts the DE density is negligible. These considerations are taken into account in our statistical analysis of the observational data. }. In Fig. 3 we display the linear growth rate $f(z_{2df})$ as a function of the index $\alpha$ for some selected values of the parameter $\tilde{A}$ and $\Omega_{m,0} = 0.27$. Note that while the quantity $f(z_{2df})$ is almost insensitive to the index $\alpha$, it depends considerably on the parameter  $\tilde{A}$.

\begin{figure}[t]
\label{fig:trans}
\centerline{\psfig{figure=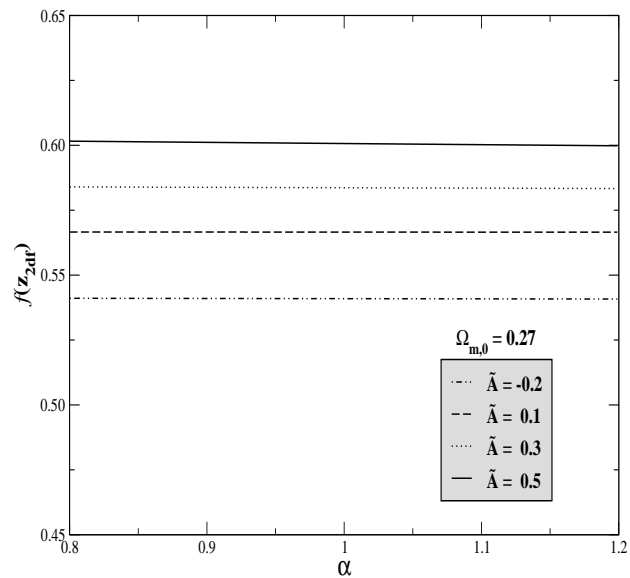,width=3.4truein,height=3.4truein,angle=-90}
\hskip 0.1in} 
\caption{The linear growth rate $f(z_{2df})$ as a function of the index $\alpha$ for some selected values of the parameter $\tilde{A}$. Note that while the quantity $f(z_{2df})$ is an insensitive function of $\alpha$, it depends considerably on the parameter  $\tilde{A}$.}
\end{figure}

In Fig. 4 we show the resulting likelihood from our statistical analysis. Contours of 68.3\%, 95.4\% and 99.73\% confidence levels (c.l.) are displayed in the $\tilde{A} - \alpha$ plane by considering a joint analysis involving the observational data described above. Note that at least two important conclusions may be drawn from this figure. First, that the allowed interval for the parameter $\tilde{A}$ is considerably restricted around the value $\tilde{A} = 0$, which is formally equivalent to the cosmological constant $\Lambda$ [see Eq. (\ref{eq:w})]. In other words, such a result amounts to saying that we cannot go very far from the vacuum EOS ($w_{eff} = -1$), in order to be consistent with the presently available cosmological observations. This is very probably true, although, the second aspect of Fig. 3 also shows that almost no constraint can be placed on the additional parameter $\alpha$ from the available sets of observational data, which hampers any definitive answer to our initial question. Such a conclusion may also be generalised for other EOS parametrizations as, for instance, the so-called Chaplygin gas EOS (Kamenshchik, Moschella and Pasquier, 2001; Bilic, Tupper and Viollier, 2002; Bento, Bertolami and Sen, 2002; Dev, Jain and Alcaniz, 2003) or still for EOS parametrizations that depend explicitely on time/redshift (Goliath et al., 2001; Efstathiou, 1999; Chevallier and Polarski, 2001; Linder, 2003; Padmanabhan and Choudhury, 2003; Jain, Alcaniz, Dev, 2006) [see also Maor et al. (2002) for a discussion on this topic]. For the analysis performed here the best-fit model is found for values of $\tilde{A} = 0.03$ and $\alpha = 0.98$, which corresponds to a 13.6-Gyr-old, currently accelerated universe with $q_0 = -0.63$. Note that 
according to our discussion presented in Section II, these values for $\tilde{A}$ and $\alpha$ also characterizes a singular fate of the universe  qualitatively similar to the ``Big Rip" singularity (Caldwell, Kamionkowski and Weinberg, 2003; Nesseris and Perivolaropoulos, 2004), in which the scale factor $a$ diverges at a finite time, as do the dark energy density $\rho_{d}$ and pressure $p_{d}$. Naturally, such a scenario cannot be taken as a prediction from the current observational data since positive and negative values of $\tilde{A}$ are possible at 1$\sigma$ level and the value of $\alpha$ is completely unconstrained by current observations. In particular, at 95.4\% c.l. the values of the parameter $\tilde{A}$ are restricted to the interval $-0.06 \lesssim \tilde{A} \lesssim 0.12$.

\begin{figure}
\vspace{.2in}
\centerline{\psfig{figure=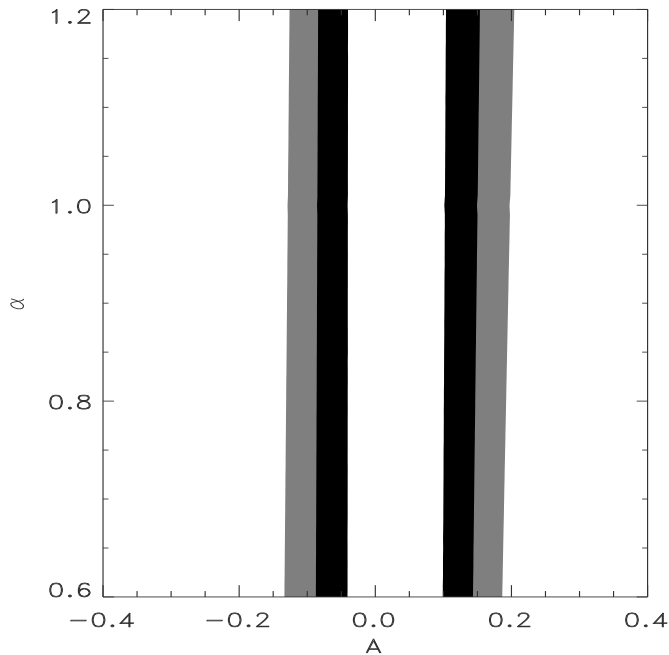,width=4.0truein,height=3.7truein,angle=0}
\hskip 0.1in} 
\caption{The plane $\tilde{A} - \alpha$ for the EOS parametrization discussed in this paper. The curves correspond to confidence regions of $68.3\%$ and 95.4$\%$ for a joint analysis
involving SNe Ia, growth of structure index and the age of the Universe, as discussed in the text.  Note that although the parameter $\tilde{A}$ is restricted to the interval $-0.06 \lesssim \tilde{A} \lesssim 0.12$ at 95.4\% c.l., the parameter $\alpha$ is completely unconstrained by the current observational data.}
\end{figure}

\section{Concluding remarks}

Although some questions about the current value of the dark energy EOS still remain to be answered, it is nowadays becoming a consensus that its effective value must be something very close to the cosmological constant prediction, i.e., $w \equiv p_d/\rho_d = -1$. However, if the functional form of the EOS of the dark component responsable for the present cosmic acceleration has additional parameters, very little can be said about their values from the available sets of observational data. This important aspect in our search for a better understanding of the physical properties of dark energy is exactly the focus of our discussion in this paper. By considering a kind of ``expansion" around the vacuum of the type $p_d = -\rho_d - A\rho_d^{\alpha}$, we have firstly discussed the rich variety of phenomena predicted for the past and future of the Universe  when different values (and/or combinations) of the parameters $\tilde{A}$ and $\alpha$ are considered. We also have studied the D/A transition and shown that for a large interval of the parameter $\tilde{A}$ the parametrization discussed here is in full agreement with the current estimate for the acceleration redshift ($z_*$), as given by Riess et al. (2004). Another interesting aspect of parametrization (1) is its prediction of models with a transient acceleration phase (D/A/D  transition), which seems to be in accordance with the requirements of String or M-theory (Fischler et al. 2001) (see also Ellis, Mavromatos \& Nanopoulos, 2005). Finally, we have also performed a statistical analysis involving some of the most recent cosmological observations to show that, while the parameter $\tilde{A}$ is tighly restricted to values around the cosmological constant  ($\tilde{A} = 0.03 \pm 0.09$ at 95.4\% c.l.), the value of $\alpha$ is completely unconstrained by the current observational data, which hampers a definitive answer to our initial question. We expect that in the near future new sets of observations along with more theoretical effort will be able to shed some light on our (so far) \emph{dark} search to measure how far we can really go from $\Lambda$.


\begin{acknowledgements}
HS acknowledges the support of the Secretar\'{\i}a de Estado de Universidades e Investigaci\'{o}n of the Ministerio de Educaci\'{o}n y Ciencia of Spain within the program ``Ayudas para movilidad de Profesores de Universidad e Investigadores espa\~{n}oles y extranjeros". He is also partially supported by MEC and FEDER under project 2004-04582-C02-01 and by the Dep. de Recerca de la Generalitat de Catalunya under contract CIRIT GC 2001SGR-00065 and would like to thank the Departament E.C.M. of the Universitat de Barcelona for the hospitality. JSA acknowledges support from Conselho Nacional de Desenvolvimento Cient\'{\i}fico e Tecnol\'ogico (Brazilian Research Agency) grants CNPq/307860/2004-3 and CNPq/475835/2004-2 and from FAPERJ No. E-26/171.251/2004.
\end{acknowledgements}

\end{document}